\documentclass[pre,aps,twocolumn,superscriptaddress]{revtex4-1}
\usepackage{latexsym,amssymb,amsfonts,amsmath,graphicx,bbm,color,bm,times,hyperref,amstext,relsize}

\newcommand{\ket}[1]{{\left\vert {#1} \right\rangle}}	
\newcommand{\bra}[1]{{\left\langle {#1} \right\vert}}
\def\braket#1{\langle{#1}\rangle}

\begin{document}

\title{Work and Quantum Phase Transitions: Is there Quantum Latency?}

\author{E. Mascarenhas}
\affiliation{Departamento de F\'isica, Universidade Federal de Minas Gerais, Belo Horizonte, MG, Brazil}

\author{H. Bragan\c{c}a}
\affiliation{Departamento de F\'isica, Universidade Federal de Minas Gerais, Belo Horizonte, MG, Brazil}

\author{R. Dorner}
\affiliation{Blackett Laboratory, Imperial College London, Prince Consort Road, London SW7 2AZ, United Kingdom}
\affiliation{Clarendon Laboratory, University of Oxford, Parks Road, Oxford OX1 3PU, United Kingdom}

\author{M. Fran\c{c}a Santos}
\affiliation{Departamento de F\'isica, Universidade Federal de Minas Gerais, Belo Horizonte, MG, Brazil}

\author{V. Vedral} 
\affiliation{Clarendon Laboratory, University of Oxford, Parks Road, Oxford OX1 3PU, United Kingdom}
\affiliation{Centre for Quantum Technologies, National University of Singapore, 3 Science Drive 2, Singapore 117543}
\affiliation{Department of Physics, National University of Singapore, 2 Science Drive 2, Singapore 117543}

\author{K. Modi}
\affiliation{School of Physics, Monash University, VIC 3800, Australia}

\author{J. Goold}
\affiliation{The Abdus Salam International Centre for Theoretical Physics (ICTP), Trieste, Italy}

\date{\today}

\begin{abstract}
We study the physics of quantum phase transitions from the perspective of non-equilibrium thermodynamics. For first order quantum phase transitions, we find that the average work done per quench in crossing the critical point is discontinuous. This leads us to introduce the quantum latent work in analogy with the classical latent heat of first order classical phase transitions. For second order quantum phase transitions the irreversible work is closely related to the fidelity susceptibility for weak sudden quenches of the system Hamiltonian. We demonstrate our ideas with numerical simulations of first, second, and infinite order phase transitions in various spin chain models.
\end{abstract}


\maketitle

\section{Introduction}

 Classical phase transitions are driven by a multitude of mechanisms such as particle or heat exchange with a reservoir~\cite{Callen}. A characteristic trait of first order classical phase transition, e.g. water turning into ice, is an exchange of heat between the system and reservoir at constant temperature called the \textit{latent heat}; this is the energy needed to go from one state of matter to another~\cite{Binder}. This can be made explicit if we consider the free energy $F=U-TS$ which has a discontinuous derivative at the first order critical point, implying a discontinuity in entropy since $S=-\frac{\partial F}{\partial T}$. Therefore we also have a discontinuity in internal energy which is the latent heat $\Delta U=T\Delta S=Q_{\mathrm{latent}}$. 
 
 Quantum phase transitions (QPTs), on the other hand, occur at zero temperature and are driven by changes in the system Hamiltonian, i.e., by extracting or performing work on the system~\cite{Sachdev}. Here, we recast QPTs in the framework of non-equilibrium thermodynamics~\cite{Tasaki, Kurchan, Mukamel, lutz,Carol} and show that the average and irreversible work can be made vanishingly small in the vicinity of a first order QPT, therefore we show that there is no correspondence with classical latency. Thus there is no quantum latent work associated with equilibrium first order quantum phase transitions. However, an actual transition between two phases separated by a first oder transition is forbidden by the hamiltonian dynamics and thus requires the presence of an external bath which allows one phase to be converted into the other. The bath absorbs the excess work as a heat transfer from the system and hence latency is found as a nonequilibrium property.

In this article we consider the work done on a quantum system when it is taken across the critical point of a QPT by an infinitesimal-instantaneous change of its Hamiltonian. The sudden quench simplifies our analysis to give a transparent interpretation of the essential physics without loss of generality. Our method relies on quantifying the non-equilibrium work by analyzing the moments of the quantum work distribution. This approach has recently been used to provide insight into both the thermodynamic and universal features of quantum many-body systems~\cite{Silva1a, Silva1b, Silva1c, dorner, Dora, Silva2,Sindona,Sindona2,Mauro,Mauro2}. In particular, it was recently shown that for a zero-temperature quantum system undergoing a sudden quench, the first and second derivatives of the ground state energy with respect to the quench parameter are closely related to the average work and irreversible work respectively~\cite{Sotiriadis}. Building on this result, we show how the work distribution captures the non-analyticity of the ground state energy in first order QPTs and to the order parameter and susceptibility of the model in second order QPTs. We support our findings with numerical simulations of the first, second, and infinite order QPTs in the $XXZ$ spin chain, which maps to a model of interacting fermions~\cite{Gia}.


\section{Pure state thermodynamics}

We consider a quantum system with the Hamiltonian ${H} (\lambda) = {H}_\textrm{free} + \lambda {V}$, where $\lambda$ is an external parameter controlling the strength of the perturbing potential ${V}$. For $t < 0$ the control parameter is held at a fixed initial value $\lambda = \lambda_\textrm{i}$ and the system is coupled to a reservoir at zero-temperature. We make two assumptions, (i) the ground state to be non-degenerate which can always be assumed for real systems in which small imperfections and disorder naturally break degeneracy and (ii) even though absolute zero temperature is not feasible it is a fair representative of currently attainable low temperature physics~\cite{Fuckyou,Cool}.
Upon equilibration, the system reaches its ground state, defined by ${H} (\lambda_\textrm{i}) \ket{\psi_0} = E_0 (\lambda_\textrm{i}) \ket{\psi_0}$ where $E_0(\lambda_\textrm{i})$ denotes the ground state energy. The control parameter is instantaneously quenched to a final value $\lambda_\textrm{f}$ giving the Hamiltonian ${H} (\lambda_\textrm{f}) = \sum_m E_m (\lambda_\textrm{f}) \ket{\phi_m} \bra{\phi_m}$ where ${E}_m (\lambda_\textrm{f})$ are the energy eigenvalues of the final Hamiltonian and $\{\ket{\phi_m}\}$ are the corresponding eigenstates.

The work done on the system is defined as the difference between the initial energy of the system and the outcome of an energy measurement performed in the eigenbasis of the final Hamiltonian, i.e., $W_m = E_m (\lambda_\textrm{f}) - E_0 (\lambda_\textrm{i})$, where the outcome $E_m (\lambda_\textrm{f})$ is obtained with probability $p_m = \vert \braket{\psi_0 \vert \phi_m} \vert^2$. Accordingly, the \textit{quantum work distribution}, which encodes the full statistics of work, is given by~\cite{Tasaki, Kurchan, Mukamel, lutz} $P(W) =\sum_m \; p_m \; \delta\left(W-W_m\right)$, which has a direct connection to the Loschmidt echo~\cite{Silva1a}.The first moment of the work distribution gives the average work done;
\begin{align}
\braket{W} =& \int W P(W) \textrm{d}W \nonumber \\
=& \bra{\psi_0} H(\lambda_\textrm{f}) \ket{\psi_0} -\bra{\psi_0} H(\lambda_\textrm{i})\ket{\psi_0} \nonumber \\
=& (\lambda_\textrm{f} - \lambda_\textrm{i}) \bra{\psi_0} {V} \; \ket{\psi_0}
= \left. \delta\lambda\frac{\partial E_0}{\partial \lambda}\right\vert_{\lambda_\textrm{i}},
\label{eq:avwork}
\end{align}
where the last equality follows from ${V}=\partial {H}/\partial \lambda$, the Hellmann-Feynman relation~\cite{Feynman}, and we define $\delta\lambda= \lambda_\textrm{f} - \lambda_\textrm{i}$.

The average work is bounded from below by the Clausius statement of the second law~\cite{Campisi}. At zero temperature, this requires that $\braket{W} \geq \Delta U$ with $\Delta U = E_0 (\lambda_\textrm{f}) -E_0 (\lambda_\textrm{i})$ denoting the change in internal energy. The Clausius inequality is saturated for completely adiabatic evolution. However, for general quenches the system can become excited, thereby dissipating work. This leads to the definition of the irreversible work $\braket {W_\textrm{irr}} = \braket{W} -\Delta U$ as a measure of the non-adiabiticity of the quench~\cite{Deffner}. For a weak quench, $\lambda_\textrm{f} -\lambda_\textrm{i}=\delta \lambda \ll 1$, the irreversible work can be expanded in powers of the small parameter $\delta \lambda$, thus,
\begin{align}
  \braket{W_\textrm{irr}}   &=\left. \delta \lambda \frac{\partial E_0}{\partial \lambda} \right\vert_{\lambda_\textrm{i}} - E_0 (\lambda_\textrm{i} +\delta \lambda) +E_0(\lambda_\textrm{i}) \nonumber \\
  &\left.\approx -\frac{\delta \lambda^2}{2}\frac{\partial^2 E_0}{\partial \lambda^2}\right\vert_{\lambda_\textrm{i}}.
  \label{eq:irrworkder}
\end{align}

\section{Universal features of QPTs}

\begin{figure}
\includegraphics[width=\linewidth]{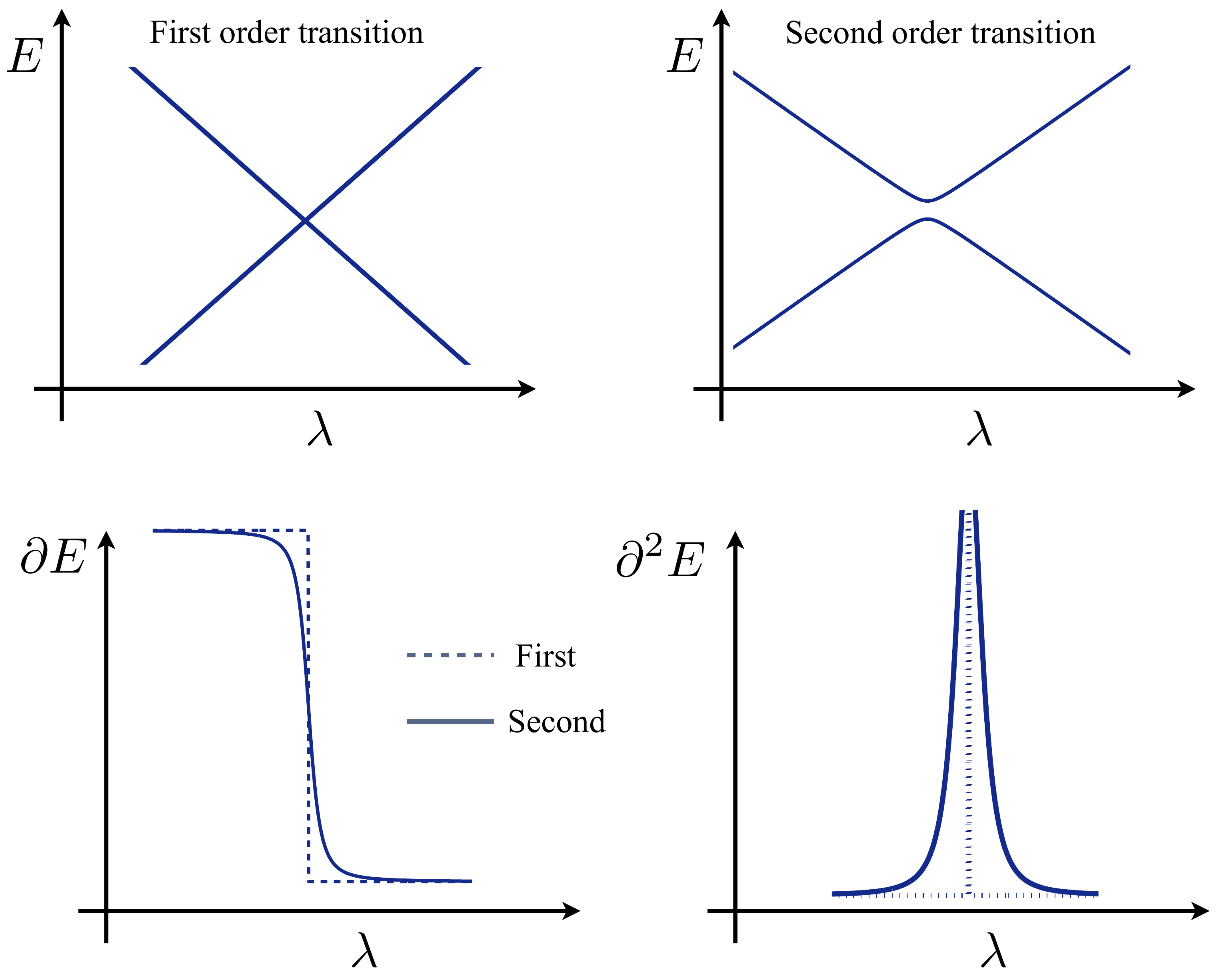}
\caption{(Color online.) Schematic representation of a level crossing giving rise to a first order QPT and an avoided crossing leading to a second order QPT as discussed in the main text. The corresponding first and second order derivatives of the ground state energy are also presented. For $\epsilon=0$, the ground and excited states of the Landau-Zener Hamiltonian (Eq.~\eqref{eq:LZ}) exhibit a level crossing as the external field is tuned through the critical value $\lambda_\textrm{c} = \Delta/(2a)$, while for $\epsilon \neq 0$, the energy levels exhibit an avoided crossing. In the case of the level crossing, both derivatives are discontinuous. For the avoided crossing, the first and second derivatives are continuous for finite $\epsilon$ and become discontinuous in the limit $\epsilon\rightarrow 0$ as the turning point in the ground state energy becomes a kink. As an aside, we note that the Landau-Zener Hamiltonian is isomorphous to the Ising mean-field Hamiltonian, which is accurate for the infinitely connected many-body lattice~\cite{Sachdev}.} \label{levelcrossing}
\end{figure}

Zero temperature quantum systems in the thermodynamic limit undergo a phase transition when a Hamiltonian parameter is tuned through a point of non-analyticity in the derivatives of the ground state energy~\cite{Sachdev}. For first order QPTs this non-analyticity takes the form of a level crossing, while for second order QPTs the critical point occurs at an avoided crossing (see Fig.~\ref{levelcrossing} for a graphical illustration). Owing to this universal behavior, we need only consider a minimal model incorporating a level crossing and an avoided crossing for our investigation of the thermodynamics of QPTs. We therefore choose the Landau-Zener model, describing a single two level system with energy splitting $\Delta$ and coupling $\epsilon$ within an externally tunable magnetic field of strength $\lambda$. The relevant Hamiltonian is
\begin{gather}
H_\textrm{LZ}= \left(-\frac{\Delta}{2}+a\lambda \right)\sigma^z+\epsilon \sigma^x,
\label{eq:LZ}
\end{gather}
where $\sigma^\alpha$ is a spin-1/2 Pauli matrix with $\alpha=\{x,y,z\}$  and $a$ measures the strength of the coupling between the two-level system and the magnetic field. 
The ground state energy of the Hamiltonian in Eq.~\eqref{eq:LZ} is easily found to be $E_0 = -\frac{1}{2} \sqrt{4 \epsilon^2+(\Delta-2a\lambda)^2}$. The first and second derivatives of the ground state energy with respect to the control parameter are then
\begin{align}
\frac{\partial E_0}{\partial \lambda} &=\frac{a (\Delta -2 a \lambda )}{\sqrt{4 \epsilon ^2+(\Delta -2 a \lambda )^2}}, \label{dacgs1}
\\
\frac{\partial^2 E_0}{\partial \lambda^2}&=-\frac{8 a^2 \epsilon ^2}{\left(4 \epsilon ^2+(\Delta -2 a \lambda )^2\right)^{3/2}}.
\label{dacgs}
\end{align}

Hence, combining Eq.~\eqref{dacgs1} with Eq.~\eqref{eq:avwork}, we are able to evaluate the average work induced by a sudden quench of the external field in the Landau-Zener model. We approximate an adiabatic change in the external parameter by considering weak sudden quenches, i.e., taking $\lambda_\textrm{i}$ to $\lambda_\textrm{f} = \lambda_\textrm{i} + \delta \lambda$. 
In the level crossing scenario ($\epsilon=0$, see Fig.~\ref{levelcrossing}), analogous to a first order QPT, the average work done per quench is, thus, $\braket{W}/\delta \lambda =  a$ for $\lambda_\textrm{f} <\lambda_\textrm{c}$ and $\braket{W}/\delta \lambda =-a$ for $\lambda_\textrm{i} > \lambda_\textrm{c}$, where  $\lambda_\textrm{c} = \Delta/(2a)$ is the critical value of the external field. Evidently, the average work per quench exhibits a discontinuity at the critical point of magnitude
\begin{gather}
w = 2a.
\end{gather}
This discontinuity is a general feature of the level crossing and, therefore, a general feature of first order QPTs. Similar discontinuous behavior is also exhibited by the classical latent heat in CPTs. This sudden change is not quantum reminiscent of classical latency rather a novel form of nonequilibrium quantum latency. 

Physically, the average amount of work required to cross the critical point of a first order QPT vanishes with the ``size'' of the quench $W=\delta \lambda w$. 
However, as the system is driven across the level crossing it inevitably becomes excited, even for very slow evolution. Hence, to bring the system to its new ground state following the quench, it must be attached to a zero-temperature reservoir. During the equilibration process, an amount of heat is dissipated from the system to the reservoir. The amount of heat transfer is given by the ``excess" energy in the system, which is exactly the irreversible contribution to the quantum work~\cite{Deffner}. Thus, we have a quantum heat per quench
\begin{gather}
q= \braket{w_\textrm{irr}}
\end{gather}
as a universal feature of first order QPTs.
However, again $Q=\delta \lambda q$ goes to zero with the size of the quench, thus, there is no heat release intrinsically associated to equilibrium first order quantum phase transitions.

For $\epsilon \neq 0$, the Landau-Zener Hamiltonian in Eq.~\eqref{eq:LZ} exhibits an avoided crossing, analogous to a second order QPT (see Fig.~\ref{levelcrossing}). Combining Eq.~\eqref{dacgs} with Eq.~\eqref{eq:irrworkder}, we are able to evaluate the irreversible work done following a weak sudden quench of the magnetic field strength. For a sudden quench beginning at the critical value of the external parameter $\lambda_\textrm{c}=\Delta/(2a)$, the irreversible work reduces to
\begin{gather}
\braket{ W_\textrm{irr}} =-\frac{\delta \lambda^2}{2} \left. \frac{\partial^2 E_0(\epsilon\neq0)}{\partial \lambda^2} \right\vert_{\lambda=\lambda_\textrm{c}}
=\frac{\delta \lambda^2 a^2}{2 \epsilon}.
\label{irrworklz}
\end{gather}
We see that as $\epsilon \rightarrow 0$ in Eq.~\eqref{irrworklz}, consistent with a second order QPT in the thermodynamic limit, the irreversible work for finite quenches at criticality diverges. This is consistent with the results of Refs.~\cite{Silva1a, Silva1b, Silva1c, dorner, Silva2, Sotiriadis} where the irreversible work is shown to indicate second order QPTs.
We also point out the similarity between the irreversible work and the fidelity susceptibility, which is also a good indicator of second order QPTs~\cite{FidelityA, FidelityB, FidelityPerturb, FidelityReview} analagous to the thermal susceptibility in a thermally driven second order CPT.

Recalling that low order transitions are associated with nonanalytical behavior in derivative of the energy of corresponding order, we have explicitly shown how this thermodynamical approach detects low order transitions. 
Having elucidated the physical lack of latency in first order QPTs and reiterated the utility of the irreversible work as a susceptibility in second order QPTs, we now proceed to demonstrate these ideas in quantum spin chains.
\begin{figure}
\includegraphics[width=\linewidth]{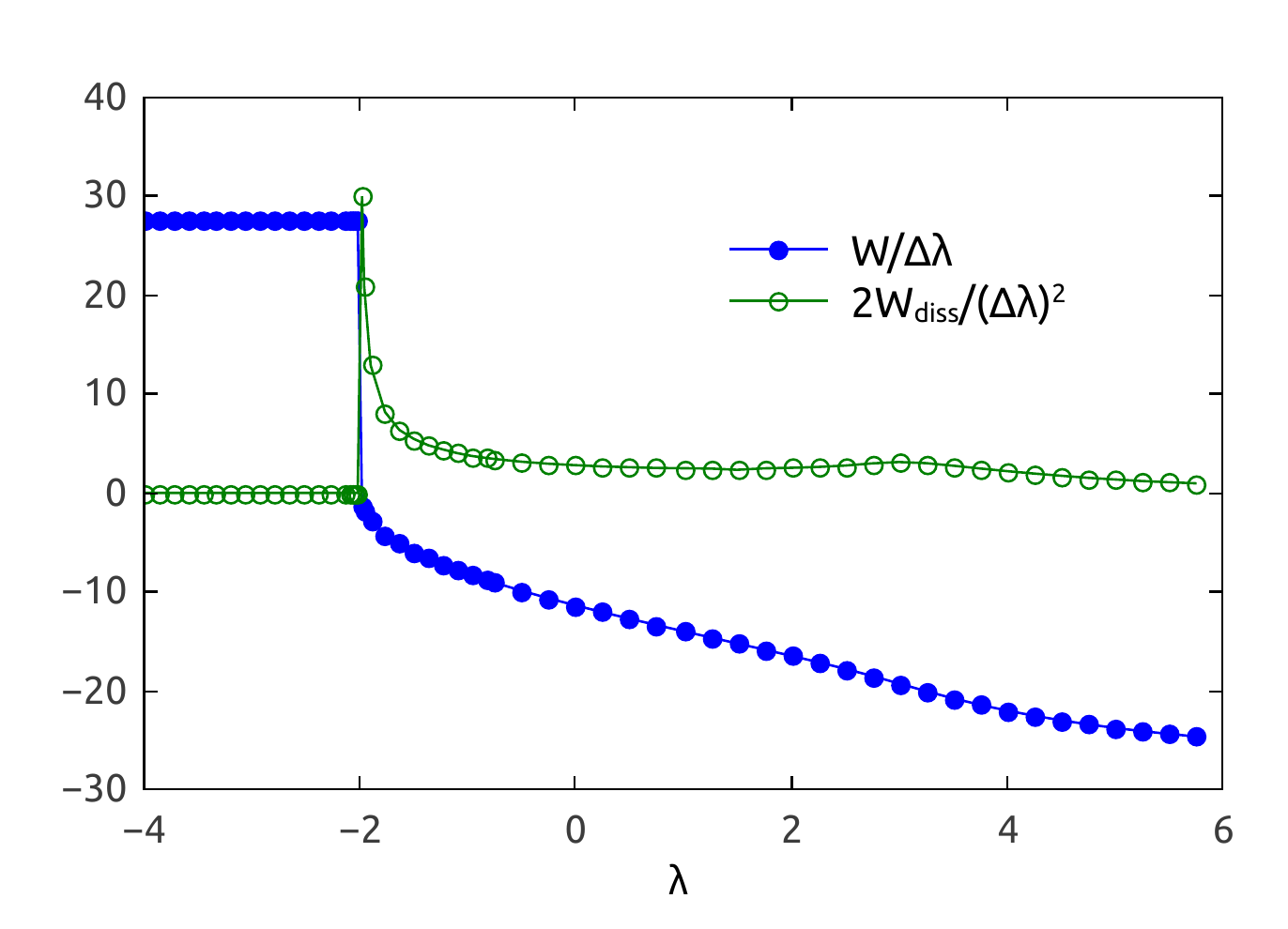}
\caption{(Color online.) Density matrix renormalization group (DMRG) data (in units of $J=1$) for the average work (filled blue circles) and irreversible work (empty green circles) done in a weak sudden quench of the $XXZ$ hamiltonian with 112 sites and $\delta\lambda=10^{-5}$. Here, we assume the presence of a small local energy shift at one lattice site which lifts the degeneracy in the ground state of the Hamiltonian. The data has a DMRG truncation error and energy convergence of $\approx10^{-9}$. The quench protocol we consider is an instantaneous change of the $Z$-coupling $\delta \lambda = \lambda_\textrm{f} - \lambda_\textrm{i} \ll 1$ with the system initialized in the ground state of the initial Hamiltonian. Both the average and irreversible work display a discontinuity at the critical point  of the first order transition at $\lambda=-2$. On the left hand side of the first order transition, the gapped spectrum of the ferromagnetic phase enforces adiabaticity as the system cannot be taken out of equilibrium by weak quenches within the same phase. This means that the system is not excited by the quench and the sole contribution to the average work is the change in internal energy. The discontinuities at the critical point indicate work required to drive the system from the ferromagnetic phase to the Luttinger liquid phase. The new phase is characterized by a continuous energy spectrum and so subsequent quenches can excite the system, leading to the dissipation of work. In contrast, neither the average work nor the irreversible work indicate the Berezinskii-Kosterlitz-Thouless transition at $\lambda=2$.}
\label{fermions}
\end{figure}

\section{Quantum many-body systems}

We choose the one-dimensional anisotropic $XYZ$ spin chain as the starting point for our investigation. This model is fully equivalent to a spin-polarized extended Hubbard model at half-filling, describing an effective system of spinless fermions~\cite{Fermi1, Fermi2}. The Hamiltonian is given by
\begin{gather} 
{H}=\sum_{\braket{ i,j}} \left[ J_x \sigma^{x}_i \sigma^{x}_j + J_y \sigma^{y}_i \sigma^{y}_j + \frac{\lambda}{2} \sigma^{z}_i \sigma^{z}_j + h\sigma^{z}_i\right],
\label{eq:fermiham}
\end{gather}
where $J_{x(y)}$ is the spin coupling along the $X(Y)$-axis, $\lambda$ is the coupling along the $Z$-axis and $h$ is the external magnetic field along the $Z$-axis. For a full discussion of the $XYZ$ model and its mapping to the fermion chain, see Ref.~\cite{Gia}.

To proceed, we consider the $XXZ$ Hamiltonian ($J_x=J_y=J$) with no external field ($h = 0$). In the parameter regime $\lambda/J<-2$ the ground state is ferromagnetically ordered (a fully-filled band insulator in the Fermi picture). A first order QPT to a Luttinger liquid phase is brought about by tuning $\lambda/J$ to the regime $\vert \lambda/J\vert <2$. At $\lambda/J=2$ the system undergoes an infinite order Berezinskii-Kosterlitz-Thouless QPT to an anti-ferromagnetic phase (a charge density wave phase in the fermionic picture). In Fig.~\ref{fermions} we show numerical results for the average work and irreversible work done following a series of weak sudden quenches across the phase diagram, passing through both critical points. 
The work and irreversible work exhibit the discontinuous behavior predicted in our analysis of the Landau-Zener model.
The origin of the discontinuity in the irreversible work, can also be explained phenomenologically in this instance; As the magnetization of the two phases is different, the dynamics induced by quenching the Hamiltonian in Eq.~\eqref{eq:fermiham}, which preserves the total magnetization, are not able to convert one phase to the other. To drive the transition, a zero-temperature reservoir must be attached to the system at the end of the quench protocol, bringing the system to its new ground state.  Physically, this corresponds to electromagnetic energy exchange between the system and the environment, which is consistent with heat exchange during a cooling process. 

For completeness, we mention that neither the work nor the irreversible work indicate the Berezinskii-Kosterlitz-Thouless transition at $\lambda/J=2$ (see Fig.~\ref{fermions}). This is expected since the Berezinskii-Kosterlitz-Thouless transition is of infinite order and is, therefore, not captured by the finite order non-equilibrium thermodynamical approach we adopt here.

\begin{figure}
\includegraphics[width=\linewidth]{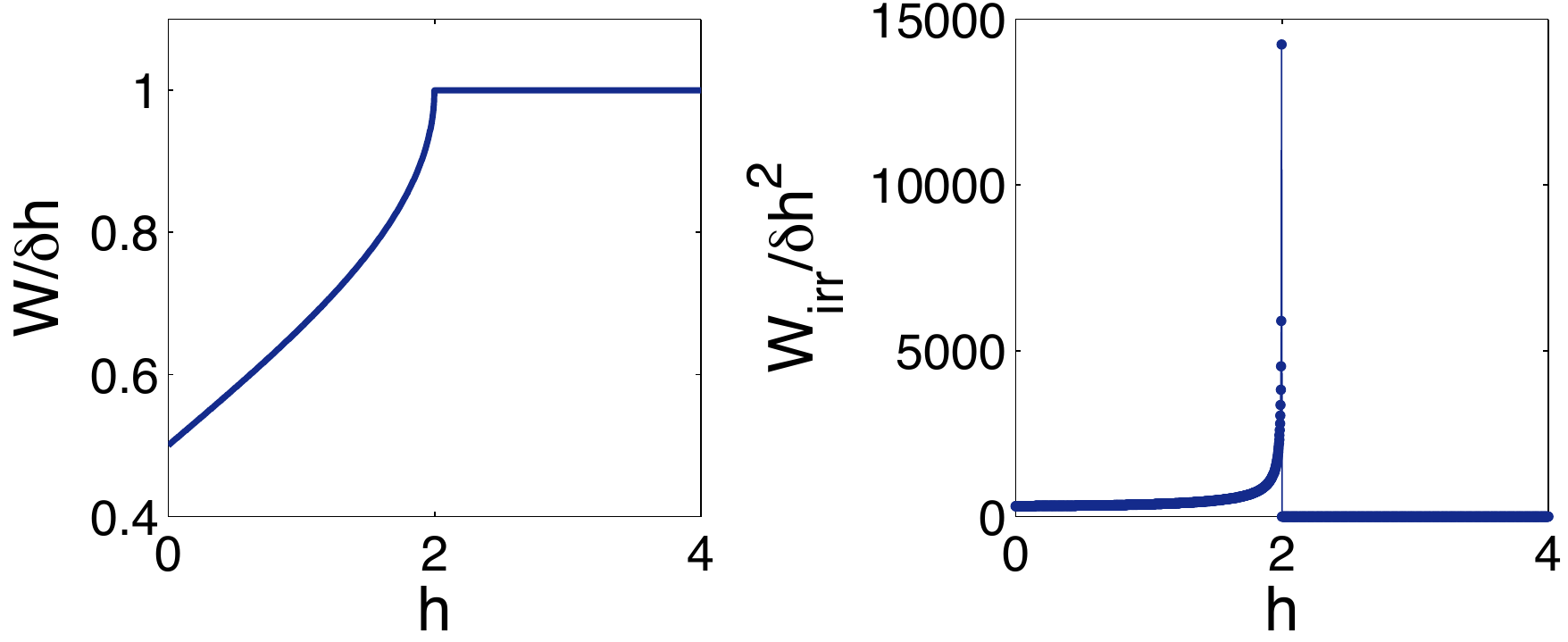}
\caption{(Color online.) Exact numerical results for the average work and irreversible work induced by a sudden quench of the external magnetic field $h$ in the $XX$ model. This model incorporates a second order QPT from a ferromagnetic phase for $h/J \ll 1$ to a paramagnetic phase for $h/J\gg1$, with the critical point occurring when the external field is two times the internal coupling $h=2J$. The phase diagram exhibits a discontinuity in the average work and a divergent irreversible work at the critical point. This behaviour reflects the relationship between the irreversible work and fidelity susceptibility discussed in the main text.}
\label{TI}
\end{figure}

We now turn our attention to second order QPTs. In Fig.~\ref{TI}, we plot the numerically exact results for the average work and irreversible work done by quenching the external field in the $XX$ model  ($\lambda=0$ and $J_x=J_y=J$ in Eq.~\eqref{eq:fermiham}). We see that the average work has a discontinuous derivative at the QPT and, thus, the irreversible work shows singular behavior at the critical point, consistent with its interpretation as a susceptibility.  The thermodynamic properties of the second order QPT in the transverse Ising model ($\lambda=J_y=0$ and $J_x=J$ in Eq.~\eqref{eq:fermiham}) have been extensively studied~\cite{Silva1a, Silva1b, Silva1c, dorner, Silva2, Sotiriadis} and shown to exhibit the same global features in the phase diagram. We also note that, in these specific cases, the average work $\braket{W} = \delta h\sum_i \braket{\sigma^z_i}$ is given by the order parameter of the model. 
The XX spin chain thermodynamical analysis seems to be connected to a geometric analysis given in~\cite{cu}. There the geometric phase and its derivative
can be written as a function of the derivatives of the ground state
energy with respect to the field strength i.e., the order parameter of
the model.

\section{Conclusion}

In this work we have analyzed the statistics of work done on general classes of quantum critical models by infinitesimal sudden quenches of a control parameter. We show that first order QPTs exhibit a discontinuity in the work distribution similar but not analogous to the classical latent heat, such that quantum latency is intrinsically a nonequilibrium phenomenon.  As a final remark, we point out that recent proposals to measure the statistics of work by means of an ancillary system~\cite{dorner2, mauro} (see also~\cite{campisihanggi} for an extension to open systems) can be extended to the many-body domain~\cite{goold:11} and used to verify our findings, hence bringing pure state thermodynamics into the laboratory.


The authors thank T. J. G. Apollaro, S. Montangero, M. C. de Oliveira Aguiar, R. Pereira, and M. Terra Cunha for helpful discussions and insightful comments. EM, HB and MFS thank CNPq (Brazil) for financial support. RD and VV are grateful for financial support from the EPSRC (UK). VV acknowledge financial support from the John Templeton Foundation, National Research Foundation and the Ministry of Education (Singapore), funding from the Leverhulme Trust and the Oxford Martin School. EM thanks University of Oxford for their hospitality and JG thanks Universidade Federal de Minas Gerais for their hospitality.

\end{document}